\documentstyle[sprocl]{article}

\newcommand{\md}{\mbox{d}}
\newcommand{\Fc}{{\cal F}}
\newcommand{\Ic}{{\cal I}}

\begin{document}

\title{INTERFERING QCD/QED VACUUM POLARIZATION}

\author{H.-TH. ELZE}

\address{Instituto de F\'{\i}sica, UFRJ, CP 68528, 
21945-970 Rio de Janeiro, Brasil} 

\author{B. M\"ULLER}

\address{Physics Department, Duke University, Durham, NC\,27708-0305, U.S.A.}

\author{J. RAFELSKI}

\address{Physics Department, University of Arizona, Tucson, AZ\,85721,
U.S.A.}

\maketitle
\abstracts{
Vacuum polarization mediated by quark loops is susceptible to external electromagnetic fields as well as to the QCD vacuum structure. Employing 
the stochastic vacuum model, we calculate the modification of the one-loop Euler-Heisenberg effective action due to stochastic color fields with the Fock-Schwinger technique. 
Our results indicate nonperturbative light quark contributions of the same order of magnitude as the usual 
QED terms. Various theoretical and experimental implications are discussed in this progress report.           
}

\section{Introduction}
The physical vacuum has become a most 
important research topic with the 
advent of quantum field theories of the fundamental interactions, of the gauge field theory of the Standard 
Model in particular.\,\cite{GT} Basic properties of matter, such as 
masses of elementary particles and ultimately the 
observed particle spectrum, are induced by the  
vacuum state and its (broken) symmetries, which in turn are determined by the underlying interactions. Therefore, it is of considerable interest
to study in any conceivable way the dynamical vacuum and, 
especially, its altogether still unknown quantum wave functional.  

Our aim here is to perform a first step in studying the interference 
between the very different color-confining QCD and charge-screening QED 
vacua. Major efforts have been launched to investigate the `melting' of the QCD vacuum and formation of a quark-gluon 
plasma in high-energy heavy-ion collisions, see Ref.\,\cite{QM} and earlier references therein. However, relatively little attention has been paid so far to the mutual influence of QCD and QED on the vacuum state and its virtual 
excitations. Naturally, since quarks carry color as well as electric charges, both contribute to vacuum polarization.
 
Following earlier theoretical studies of the vacuum induced
nonlinear photon-photon interaction,\,\cite{BB70} there are
ongoing searches for the QED vacuum birefringence effect in high-precision crossed laser/magnetic-field experiments.\,\cite{EXP}  Most recently, we proposed to evaluate the influence of the QCD vacuum structure on the 
Euler-Heisenberg (EH) effective action of QED.\,\cite{EH35,JR98}  
This well-known one-loop effective action has been the central quantity 
for the evaluation of vacuum polarization 
effects in external electromagnetic fields, most 
notably Schwinger's calculation of pair production in 
a constant electric field.\,\cite{Schw51}   
   
Our intention is to calculate the nontrivial correction to the EH effective action arising from QCD interactions and the QCD vacuum, in particular. In lack of a solution of the QCD vacuum problem from first principles, we resort to the {\it s}tochastic {\it v}acuum {\it m}odel (SVM),\,\cite{SVM} which successfully describes color confinement, the quarkonia phenomenology, and produces interesting results in high-energy scattering, see Ref.\,\cite{DFK} and references therein. This model allows us to calculate 
the one-loop QCD vacuum-to-vacuum amplitude in the  
presence of external electromagnetic fields. 

By means of the SVM, the 
action-weighted functional integration over gluon 
(gauge potential) fields,  
$B_\mu^a,\,a=1,\dots,N_c^2-1$ for $SU(N_c)$, is replaced by an ensemble average over a Gaussian distribution of the color fields, $G_{\mu\nu}^a$, instead. The latter is characterized by a single nonvanishing vacuum correlation function, which has been computed in nonperturbative lattice gauge theory \cite{LGT} and is related to the vacuum correlator 
appearing in QCD sum rules.\,\cite{SR}  
   
We apply the Fock-Schwinger technique in order to 
calculate the one-loop effective action of 
quarks interacting nonperturbatively with an ensemble of  time 
independent homogeneous color fields as well as with an approximately constant electromagnetic background field.

Similarly as the colormagnetic instability of the Mantinyan-Savvidy vacuum ansatz,\,\cite{Savv77} the SVM points to the importance of inhomogeneous and presumably time dependent color fields, which produce a characteristic correlation length for the QCD vacuum. By the uncertainty principle, 
a typical vacuum polarization loop has an extent of 
$l\approx\Delta\tau\cdot c\approx\hbar c/(2mc^2)$, which amounts to 
$l_e\approx 200\,\mbox{fm}$ for electrons and $l_u\approx 20\,\mbox{fm}$ 
for the lightest up-quarks. Of course, the physically 
relevant value of $l_u$ must be expected to differ  considerably from this estimate guided by perturbation 
theory. Results from lattice calculations indicate that   important nonperturbative corrections have to be  
taken into account. One may consider an estimate of a lower bound to be determined by a typical constituent quark 
mass of $m_Q\approx 300\,\mbox{MeV}$ instead, i.e. 
$l_Q\approx 0.3\,\mbox{fm}$.       

Whereas all macroscopic (laboratory) electromagnetic fields are constant on the scale of $l_e$, the 
QCD vacuum correlation length obtained from lattice calculations, $\lambda\approx 0.2\dots 0.4\,\mbox{fm}$, 
is of the same order of magnitude as $l_Q$ and, therefore,   
is expected to remain small on the scale of the light quark loops.\,\cite{LGT} 

Therefore, truly space-time dependent color fields also have to be incorporated 
in the future. However, their study needs a technically different 
approach still to be developed. 
 
\section{One-loop QCD/QED Effective Action} 

We briefly rederive the one-loop effective action for fermions 
in color plus electromagnetic background fields, aiming at the case 
of a constant electromagnetic field strength with stochastic gluon 
field fluctuations, in particular. -- The calculations are performed for Minkowski space 
with the metric $g_{\mu\nu}=(1,-1,-1,-1)$, any 
Lorentz or color indices occuring 
twice are to be summed over, and we choose units such that $\hbar =c=1$. 
 
We are interested in the vacuum-to-vacuum amplitude which 
determines the QCD effective action, $\Gamma_A$, in the presence of an electromagnetic 
background potential, $A_\mu$: 
\begin{equation} \label{Gamma} 
\exp (i\Gamma_A) = \int{\cal D}\bar\psi{\cal D}\psi{\cal D}B\;
\exp\left\{ i\int\md^4x\,(L_{A,B}-\frac{1}{4}F^2-\frac{1}{4}G^2)\right\}
\;\;, \end{equation} 
where $L$ denotes the Dirac Lagrangian for the quarks coupled to the 
electromagnetic and color gauge fields, $A$ and $B$, respectively, 
$L_{A,B}\equiv\bar\psi (i\gamma\cdot D-m)\psi$, with 
$D\equiv\partial -ieA-igB$ denoting the covariant derivative in 
the fundamental representation of the color $SU(N_c)$ group; 
eventually, one has to sum over the contributions of all fermions, i.e. with different charges and masses. The gauge field actions are written in terms of 
the field strength tensors, $F_{\mu\nu}\equiv\partial_\mu A_\nu
-\partial_\nu A_\mu$ and $G_{\mu\nu}^a\equiv\partial_\mu B_\nu^a
-\partial_\nu B_\mu^a+gf_{abc}B_\mu^bB_\nu^c$, respectively. We did 
not add gauge fixing terms, since the gluon gauge potentials  
will be treated in terms of stochastic background field strengths in this work. We remark that the vacuum 
contribution to $\Gamma_A$ is selected by applying the Feynman 
boundary condition, which will be made explicit shortly.

In order to proceed, we define the Gaussian stochastic ensemble and its  correlators,\,\cite{SVM} 
which will be employed in the following evaluation of the effective action. 
We assume for simplicity that only approximately constant colormagnetic fields contribute which, however, fluctuate in 
direction and amplitude. 

As previously pointed out in Refs.\,\cite{CY80}  
in the context of QCD background field calculations, the commutator of two {\it covariantly constant} color field strength tensors vanishes, 
$[G_{\mu\nu},G_{\rho\sigma}]=0$\,. Consequently, they can be parametrized 
as being proportional to the generators of the abelian Cartan (sub-)algebra of the $SU(N_c)$ color group. 

We thus take into account $SU(3)$ colormagnetic fields of the form 
$\vec B\equiv\frac{1}{2}(\vec B^3\lambda_3+\vec B^8\lambda_8)$, involving  
the diagonal Gell-Mann matrices $\lambda_{3,8}$. Then, their normalized probability 
distribution is defined by: 
\begin{equation} \label{distrib} 
P(g\vec B^a)\;\md^3(gB^a)
\equiv (\frac{3}
{2\pi\langle g^2\vec B_a^{\;2}\rangle_G})^{3/2}
\exp (\frac{3(g\vec B^a)^2}
{2\langle g^2\vec B_a^{\;2}\rangle_G})\;
\md^3(gB^a)
\;\;, \end{equation}
for $a=3,8$\,, with the width determined by the relevant correlator.
 
In accordance with the SVM model,\,\cite{SVM} we replace the functional 
integral $\int {\cal D}B$ in Eq.\,(\ref{Gamma}) by the Gaussian ensemble average with the distribution of Eq.\,(\ref{distrib}), which we denote 
by $\langle\dots\rangle_G$ from now on.     

In the absence of a calculation from first principles, the correlator 
in Minkowski space is obtained by analytical continuation 
of the Fourier transform of the euclidean one, which is calculated numerically in lattice studies.\,\cite{LGT} This procedure is 
discussed in detail in Ref.\,\cite{DFK}. Since we consider 
only stochastic colormagnetic fields, we identify the 
widths of their distributions with the value of the 
SVM field correlator: 
\begin{equation} \label{correlator} 
\langle\frac{g^2}{2}G_{\mu\nu}^aG_a^{\mu\nu}\rangle_G
\equiv\langle g^2\vec{B}^3\cdot\vec{B}^3\rangle_G     
+\langle g^2\vec{B}^8\cdot\vec{B}^8\rangle_G
=2\langle g^2\vec{B}^3\cdot\vec{B}^3\rangle_G
\;\;, \end{equation} 
where $g$ denotes the renormalized QCD coupling. The last equality (isotropy in color space) allows us to relate the respective correlators 
to the physical gluon condensate:\,\cite{SVM} 
\begin{equation} \label{condensate}
\langle\frac{\alpha_s}{\pi}G_{\mu\nu}^aG_a^{\mu\nu}\rangle_G
=\langle\frac{\alpha_s}{\pi}:G_{\mu\nu}^aG_a^{\mu\nu}:\rangle
=0.024\pm 0.011\,\mbox{GeV}^4
\;\;, \end{equation} 
where the (running) strong coupling constant is $\alpha_s\equiv g^2/4\pi$, 
and we cite the empirical value of the condensate 
($\approx (394\,\mbox{MeV})^4$).\,\cite{LGT,SR,Nar}

For a selfconsistent determination 
of the stochastic field ensemble, the vacuum condensate 
in particular, the gluonic contribution to the effective action has to be calculated incorporating gluon fluctuations in the stochastic background together with 
an appropriate gauge fixing.\,\cite{Abb81} Here we limit our attention to the fermionic contribution, which generalizes the familiar  
Euler-Heisenberg effective action of QED.  

Obviously,  
the usual EH effective action \cite{EH35,Schw51} for leptons, $\Gamma_{\mbox{EH}}$, is obtained from Eq.\,(\ref{Gamma}) by dropping completely 
the integration over the    
gluon field $B$ and setting the corresponding terms of the classical 
action to zero. In order to evaluate $\Gamma_A$ instead, employing the SVM model with the stochastic ensemble average defined in Eqs.\,(\ref{distrib})--(\ref{condensate}), we begin by     
integrating out the fermions: 
\begin{equation} \label{Gamma1}
i\Gamma_A^- = \ln\left (\left\langle
\det\{[\gamma\cdot D+im][\gamma\cdot\partial +im]^{-1}\}\right\rangle_G
\right )
\;\;, \end{equation}
where we subtracted the contribution of the noninteracting fermionic vacuum fluctuations 
for later convenience. From now on, the Feynman boundary condition is 
implemented by replacing $m\longrightarrow m-i\epsilon$, which we often 
suppress. 

The above result, Eq.\,(\ref{Gamma1}), is similar to the usual QCD 
effective action (for $A_\mu =0$), however, with the vacuum correlator 
appearing implicitly instead of the gluon propagator. 
Next, interchanging the order of taking   
the logarithm and the ensemble average and using  
$\det\{ M\}=\exp\mbox{Tr}\ln\{ M\}$, the result is:    
\begin{equation} \label{Gamma2}
i(\Gamma_A^-)_{1-\mbox{loop}} = \left\langle\mbox{Tr}\ln 
\{[\gamma\cdot D+im][\gamma\cdot\partial +im]^{-1}\}\right\rangle_G
\;\;, \end{equation}
with a trace over space, spin, and color. We remark that interchanging
the logarithm with the ensemble average results in the 1-loop approximation 
here (suppressing this subscript henceforth). 
Since the dynamics of the Dirac field does not depend on the sign of the 
mass term, taking the average of both possibilities, the 1-loop action can be expressed conveniently, 
\begin{eqnarray} \label{Gamma3}
i\Gamma_A^-&=& \frac{1}{2}\left\langle\mbox{Tr}\ln 
\{[(\gamma\cdot D)^2+m^2]
[(\gamma\cdot\partial )^2+m^2]^{-1}\}\right\rangle_G
\nonumber \\ [1ex]
&=&
\frac{1}{2}\left\langle\mbox{Tr}\ln 
\{[\Pi^2-m^2 \right.
+\frac{1}{2}\sigma_{\mu\nu}(eF^{\mu\nu} \left. +gG^{\mu\nu})]
[P^2-m^2]^{-1}\}\right\rangle_G
,\; \end{eqnarray}
where we introduced the kinetic and canonical momentum operators, 
$\Pi_\mu\equiv iD_\mu =P_\mu +eA_\mu +gB_\mu$ and $P_\mu\equiv i\partial_\mu$, respectively; the second equality follows with the 
help of $\{\gamma_\mu ,\gamma_\nu\}=2g_{\mu\nu}$
and $[\gamma_\mu ,\gamma_\nu ]\equiv -2i\sigma_{\mu\nu}$, i.e.
the (anti-) commutation relations of $\gamma$-matrices.      
  
Finally, using the integral representation of the logarithm which 
presents the starting point of the Fock-Schwinger proper time method, 
we obtain the familiar looking result:\,\cite{Schw51}  
\begin{eqnarray} \label{Gamma4}
\Gamma_A^-=
\int_0^\infty\frac{\md s}{2is}\left\langle\mbox{Tr} 
\{\exp (is[\Pi^2-m^2 \right.
+\frac{1}{2}\sigma_{\mu\nu}(eF^{\mu\nu}+gG^{\mu\nu})])
\nonumber \\
\left.
-\exp (is[P^2-m^2])\}\right\rangle_G
\;\;, \end{eqnarray}
with the mass ($\rightarrow m-i\epsilon$) incorporating the Feynman boundary condition. 
  
\section{Fock-Schwinger Technique}  
  
For arbitrarily varying background fields it is unknown 
how to evaluate  
Eq.\,(\ref{Gamma4}), even for the QED case. Presently, we work with the simplified ensemble of color background fields introduced in the previous section, which have a fluctuating amplitude with respect to space and (color) space direction but are covariantly constant otherwise. We also incorporate the approximately constant electromagnetic field to all orders, expanding the results only in 
the end. 

We proceed by relating the first exponential in Eq.\,(\ref{Gamma4}) to an unitary evolution operator, $U(s)$, depending on the proper time variable $s$: 
\begin{equation} \label{UopH}  
U(s)\equiv\exp (-isH)\;\;, \;\;\; 
H\equiv\Pi^2-m^2+\frac{1}{2}\sigma_{\mu\nu}(eF^{\mu\nu}+gG^{\mu\nu})
\;\;, \end{equation} 
where $H$ plays the role of a fictitious Hamiltonian. 
Following the Fock-Schwinger strategy as described in Ref.\,\cite{Schw51,IZ}, the aim is to obtain the coordinate space matrix elements of $H$. Using these, the equation of motion for the evolution operator can be converted into an ordinary differential 
equation. Its solution provides the coordinate space matrix elements 
of $U(s)$, which are sufficient to calculate the trace and stochastic average in 
Eq.\,(\ref{Gamma4}). 

To begin with, introducing the 
Heisenberg operators $\Pi (s)=U^\dagger (s)\Pi U(s)$ and 
$x(s)=U^\dagger (s)x U(s)$, we obtain the related Ehrenfest 
equations of motion: 
\begin{eqnarray} \label{xofs}
\partial_sx_\mu =i[H,x_\mu ]&=&-2\Pi_\mu
\;\;, \\ [1ex]   
\label{piofs}
\partial_s\Pi_\mu =i[H,\Pi_\mu ]&=&
2(eF_{\mu\lambda}+gG_{\mu\lambda})\Pi^\lambda +igJ_\mu
+\frac{1}{2}g[D_\mu ,\sigma\cdot G] 
\\
\label{piofsa}  
&\approx&2(eF_{\mu\lambda}+gG_{\mu\lambda})\Pi^\lambda
\;\;, \end{eqnarray} 
where the color current, $J_\mu\equiv [D^\lambda ,G_{\mu\lambda}]$, 
is introduced to indicate the physical meaning of this term; similarly,  
the last term contributes the spin-color coupling here, with  
$\sigma\cdot G\equiv\sigma_{\mu\nu}G^{\mu\nu}$,
while the first term is related to the electromagnetic and color Lorentz forces. In order to arrive at Eqs.\,(\ref{xofs}) and (\ref{piofs}), we made use of the coordinate representation of the operators $\Pi ,\, x$ introduced 
after Eq.\,(\ref{Gamma3}) and treated $F_{\mu\nu}$ as constant. 
The last equality, Eq.\,(\ref{piofsa}), presents the approximation 
for covariantly constant fields, i.e. our starting point here. 
  
It is convenient to introduce a combined field strength 
tensor,
\begin{equation} \label{superF} 
(\Fc )_{\mu\nu}^{ij}\equiv eF_{\mu\nu}\delta^{ij}+gG_{\mu\nu}^at_a^{ij}
\;\;, \end{equation}
which is a matrix in Lorentz and color indices,  
with $t_a,\,a=1,\dots N_c^2-1$ denoting the generators of $SU(N_c)$ in the fundamental representation. Then, the solutions of Eqs.\,(\ref{xofs}) and (\ref{piofsa}) are easily obtained:
\begin{eqnarray} \label{pizero}
\Pi(s)&=&\exp (2\Fc s)\Pi (0)=\frac{\Fc}{\exp (-2\Fc s)-1}[x(s)-x(0)]
\;\;, \\ [2ex]
\label{xzero}
x(s)&=&\frac{1-\exp (2\Fc s)}{\Fc}\Pi (0)+x(0) 
\;\;, \end{eqnarray} 
where we employed Eq.\,(\ref{xzero}) to 
eliminate $\Pi (0)$ from Eq.\,(\ref{pizero}). 

With the help of Eq.\,(\ref{xzero}) and the basic commutator
$[\Pi_\mu (0),x_\nu (0)]=ig_{\mu\nu}$ one derives: 
\begin{equation} \label{xcommutator}
\left [x_\mu (\tau ),x_\nu (\tau ')\right ]=-2i\left (
\exp (\Fc [\tau -\tau '])
\frac{\sinh (\Fc [\tau -\tau '])}{\Fc}\right )_{\mu\nu} 
\;\;. \end{equation} 
Using this, the Hamiltonian of Eqs.\,(\ref{UopH}) can be written 
in time-ordered form and, then, its coordinate space matrix elements  
evaluated.\,\cite{Schw51,IZ} We obtain:
\begin{equation} \label{Hmatrix}
H(x',x;s)=
\frac{1}{4}[x'-x]\frac{\Fc ^2}{\sinh ^2(\Fc s)}[x'-x]
-\frac{i}{2}\mbox{tr}_L\left [\Fc\coth (\Fc s)\right ]
-m^2+\frac{1}{2}\sigma\cdot\Fc 
\;\;, \end{equation} 
where the trace refers to the Lorentz indices.

Next, we turn to the equation of motion 
for the proper time evolution operator $U(s)$, which follows from Eq.\,(\ref{UopH}). Interchanging between the Schr\"odinger 
and Heisenberg picture, we obtain in the coordinate 
representation: 
\begin{eqnarray} \label{UopHeq}
\partial_sU(x',x;s)&\equiv &
\partial_s\langle x'|U(s)|x\rangle = 
\partial_s\langle x'(s)|x(0)\rangle = 
-i\langle x'|H\;U(s)|x\rangle
\nonumber \\ [1ex]
&=&-i\langle x'(s)|H(s)|x(0)\rangle =
-iH(x'(s),x(0);s)\langle x'(s)|x(0)\rangle
\nonumber \\ [1ex]
&\equiv&-iH(x',x;s)U(x',x;s) 
\;\;, \end{eqnarray} 
with $H(x',x;s)$ from Eq.\,(\ref{Hmatrix}).         
This presents an ordinary first order differential equation for $U(x',x;s)$ as a function of $s$, which  
can be integrated directly. Since all matrices involved are  
considered as constants at this point, no ordering prescription for 
the resulting exponential is needed:
\begin{equation} \label{Uresult} 
U(x',x;s)=C(x',x)
\exp (-i\int^s\md\tau\; H(x',x;\tau ))    
\;\;. \end{equation}  
The function $C(x',x)$ incorporates the usual QED  
Schwinger string,\,\cite{Schw51,IZ} 
\begin{equation} \label{link}
C(x',x)\equiv C\exp (ie\int_x^{x'}\md z^\mu A_\mu (z))
\;,\;\; z(\xi )\equiv x+\xi (y-x)\;,\;\; 0\le\xi\le 1
\;\;, \end{equation}
as well as a normalization constant $C$. Furthermore, 
we calculate: 
\begin{eqnarray} \label{Hexp} 
-i\int^s\md\tau\; H(x',x;\tau )&=&
\frac{i}{4}[x'-x]\Fc\coth (\Fc s)[x'-x]
-\frac{1}{2}\mbox{tr}_L[\ln (\frac{\sinh (\Fc s)}{\Fc s})]
\nonumber \\
&\;&+\ln (s^{-2})
+im^2s-\frac{i}{2}\sigma\cdot\Fc s 
\;\;, \end{eqnarray} 
where we separated the second logarithm for later convenience. 

We remark that the Hamiltonian, Eq.\,(\ref{Hmatrix}), is covariant 
under the global $SU(N_c)$ gauge transformations admitted here and invariant under arbitrary  
electromagnetic gauge transformations. Therefore, the evolution operator,
Eq.\,(\ref{Uresult}), requires the additional electromagnetic string 
factor, in order to be properly covariant under either gauge transformations.  
 
Finally, the normalization constant $C$ takes the 
boundary condition (orthogonality and normalization of coordinate eigenstates) into account, 
\begin{equation} \label{normalization}
\lim_{s\rightarrow 0}U(x',x;s)=\lim_{s\rightarrow 0}\langle x'(s)|x(0)\rangle = \delta^4(x'-x)\cdot 1_{S,C} 
\;\;, \end{equation} 
with a unit matrix for spin and color on the right-hand side. The 
normalization constant is calculated using 
Eqs.\,(\ref{Hmatrix})--(\ref{normalization}). The result is: 
\begin{equation} \label{normconstant}
C=\frac{-i}{(4\pi )^2}
\;\;. \end{equation} 
Furthermore, Feynman's $m\longrightarrow m-i\epsilon$ provides the convergence factor  
in Eq.\,(\ref{Uresult}), thus implementing the asymptotic condition 
$U(x',x;s\rightarrow -\infty )=0$.
  
This completes our derivation of the (matrix elements of the) evolution operator $U$. We note in passing that these results immediately 
yield the propagator in the considered combination of  
color and electromagnetic background fields,\,\cite{Schw51,IZ} 
which may be useful for other purposes. 
  
\section{QCD-modified Euler-Heisenberg Lagrangian}    
 
Here we employ our assumptions about the nature of the stochastic color fields, in order to calculate the effective action, Eq.\,(\ref{Gamma4}), 
using the results of the previous section,
Eqs.\,(\ref{Uresult})--(\ref{normconstant}) in particular. 
 
We begin by evaluating the traces over spin and space in: 
\begin{equation} \label{traceSxU}
\mbox{tr}_{S,x}U^\dagger (s)=
\frac{+i}{(4\pi s)^2}
\int\md^4x\;\mbox{tr}_S\{\exp (
-\frac{1}{2}\mbox{tr}_L[\ln (\frac{\sinh (\Fc s)}{\Fc s})]
-im^2s+\frac{i}{2}\sigma\cdot\Fc s )\} 
\;. \end{equation}
We recall that the covariantly constant color fields are  parametrized proportional to the generators of the abelian subgroup of $SU(N_c)$, as discussed in the context of Eqs.\,(\ref{distrib})--(\ref{condensate}), i.e. $\Fc$ is a 
diagonal color matrix here. Then, the trace over spin 
is calculated similarly as in the QED case:\,\cite{Schw51}
\begin{equation} \label{traceS} 
\mbox{tr}_S\exp (\frac{i}{2}\sigma\cdot\Fc s)
=2(\cos [\frac{s}{2}\sqrt{\Ic_1+i\Ic_2}]
+\cos [\frac{s}{2}\sqrt{\Ic_1-i\Ic_2}])
\;\;, \end{equation}
where two Lorentz invariants of the fields enter, 
\begin{equation} \label{Ic1Ic2}
\Ic_1\equiv 2\Fc_{\mu\nu}\Fc^{\mu\nu}\;\;,\;\;\;
\Ic_2\equiv\epsilon_{\alpha\beta\gamma\delta}
\Fc^{\alpha\beta}\Fc^{\gamma\delta}
\;\;. \end{equation} 
The remaining 
exponential factor in Eq.\,(\ref{traceSxU}) can also be simplified: 
\begin{equation} \label{traceL}
\exp (-\frac{1}{2}\mbox{tr}_L[\ln (\frac{\sinh (\Fc s)}{\Fc s})])
=\frac{-is^2}{4}\Ic_2 (
\cos [\frac{s}{2}\sqrt{\Ic_1+i\Ic_2}]
-\cos [\frac{s}{2}\sqrt{\Ic_1-i\Ic_2}] )^{-1}
, \end{equation}
again quite similarly as in the QED case.\,\cite{Schw51} 

Combining Eqs.\,(\ref{traceSxU})--(\ref{traceL}) and using these in 
Eq.\,(\ref{Gamma4}), we obtain the unrenormalized 1-loop effective Lagrangian ($\Gamma_A^-=\int\md^4x\;{\cal L}_A^-$): 
\begin{eqnarray} \label{Gammafin}
{\cal L}_A^-&=&(-1/8\pi^2)\cdot
\\
\nonumber
&\phantom .&
\int_0^\infty\frac{\md s}{s^3}\mbox{e}^{-im^2s}
\langle\mbox{tr}_C\{
\frac{is^2}{8}\Ic_2\frac{
\cos [\frac{s}{2}\sqrt{\Ic_1+i\Ic_2}]
+\cos [\frac{s}{2}\sqrt{\Ic_1-i\Ic_2}]}{
\cos [\frac{s}{2}\sqrt{\Ic_1+i\Ic_2}]
-\cos [\frac{s}{2}\sqrt{\Ic_1-i\Ic_2}]}-1_C
\}\rangle_G
\;, \end{eqnarray}
where the color trace and the stochastic average are still left to be done.
If we omit these and set all color fields to zero, the usual 
QED result is recovered.\,\cite{Schw51,IZ}

\subsection{Renormalization}

Since QCD and QED are both renormalizable, we are 
guaranteed that the ultraviolet ($s\rightarrow 0$) divergences contained 
in ${\cal L}_A^-$ can be absorbed by renormalization, of the charges and 
fields in particular. At present we are mostly interested to demonstrate how QCD modifies the usual Euler-Heisenberg  effective Lagrangian. Therefore, it is convenient to subtract from ${\cal L}_A^-$ the pure QCD 
contribution, ${\cal L}_0^-$. 

It is calculated by setting 
$e\equiv 0$ in Eq.\,(\ref{Gammafin}). Then, for colormagnetic vacuum fields, we obtain the renormalized result: 
\begin{equation} \label{LQCD}
{\cal L}_0^-=
-\frac{1}{8\pi^2}\langle\;\mbox{tr}_C\{
g^2\vec B\cdot\vec B\int_0^\infty\frac{\md x}{x^3}\mbox{e}^{-zx}
[x\coth (x)-1-\frac{1}{3}x^2]\}\;\rangle_G 
\;\;, \end{equation}
where $z\equiv m^2/\sqrt{g^2\vec B\cdot\vec B}$ is a diagonal color 
matrix. We remark that here we rotated the integration $s\rightarrow -ix$ 
in the complex plane, as compared to Eq.\,(\ref{Gammafin});   
it is obvious now that the integral increases strongly with 
decreasing $m$. The corresponding QED integral has been calculated 
analytically.\,\cite{DittReu84} Using this, we obtain in the limit of strong fields ($z\ll 1$):
\begin{equation} \label{LQCDstrong} 
{\cal L}_0^-=\frac{-1}{24\pi^2}\langle\;\mbox{tr}_C\{
g^2\vec B\cdot\vec B[\ln (m^2/\sqrt{g^2\vec B\cdot\vec B})+\mbox{O}(z^0)]  
\}\;\rangle_G 
\;\;, \end{equation}
which is the appropriate limit for the light quarks, given the 
large value of the gluon condensate, cf. Eq.\,(\ref{condensate}). 
 
After the QCD renormalization, we have to replace $gG_{\mu\nu}=g_RG_{\mu\nu ,R}$ also in the remaining terms, i.e. 
by the renormalized quantities; we drop the subscript $R$ henceforth.  
The resulting {\it QCD-subtracted} effective Lagrangian   
is: ${\cal L}_A\equiv {\cal L}_A^--{\cal L}_0^-$, cf. 
Eqs.\,(\ref{Gammafin})--(\ref{LQCDstrong}). 
 
In order to proceed, we introduce some useful abbreviations:
\begin{eqnarray} \label{ad}
&\phantom .&\tilde a\equiv\frac{1}{4}(2F_{\mu\nu}F^{\mu\nu}
-i\epsilon_{\alpha\beta\gamma\delta}F^{\alpha\beta}F^{\gamma\delta})
\;\;,\;\;\; \tilde d\equiv \tilde a^\ast
\;\;, \\ \label{be}
&\phantom .&\tilde b\equiv \frac{g}{2}(2F_{\mu\nu}G^{\mu\nu}
-i\epsilon_{\alpha\beta\gamma\delta}F^{\alpha\beta}G^{\gamma\delta})
\;\;,\;\;\; \tilde e\equiv \tilde b^\ast
\;\;, \\ \label{cf}
&\phantom .&\tilde c\equiv \frac{g^2}{4}(2G_{\mu\nu}G^{\mu\nu}
-i\epsilon_{\alpha\beta\gamma\delta}G^{\alpha\beta}G^{\gamma\delta})
\;\;,\;\;\; \tilde f\equiv \tilde c^\ast
\;\;, \end{eqnarray} 
where $F_{\mu\nu}$ and $G_{\mu\nu}$ are understood as a unit 
and a diagonal color matrix, respectively.  
Then, the field dependent factor in the integrand of ${\cal L}_A^-$, Eq.\,(\ref{Gammafin}), is expressed as: 
$$\frac{
\cos [s\sqrt{\tilde ae^2+\tilde be+\tilde c}]
+\cos [s\sqrt{\tilde de^2+\tilde ee+\tilde f}]}{
\cos [s\sqrt{\tilde ae^2+\tilde be+\tilde c}]
-\cos [s\sqrt{\tilde de^2+\tilde ee+\tilde f}]}
\left ((\tilde a-\tilde d)e^2+(\tilde b-\tilde e)e 
+\tilde c-\tilde f\right ),$$
where $\tilde c=\tilde f$, if we assume only colormagnetic vacuum 
fields. For the following, we make use of symbolic calculations with  
{\it Mathematica},\,\cite{Wolf} of Taylor expansions in powers of $e$ in particular, for which this rewriting helps to organize 
and cut down the size of the lengthy expressions.   

Thus, the ultraviolet ($s\rightarrow 0$) structure of the {\it unrenormalized}  
but QCD-subtracted effective Lagrangian emerges after expanding up to and including terms of $O(e^3)$: 
\begin{eqnarray} \label{LQEDdiv} 
{\cal L}_A&=&\frac{1}{8\pi^2}e^2\int_0^\infty\frac{\md s}{s^3}
\mbox{e}^{-im^2s}
\\ \nonumber 
&\phantom .&\cdot\langle\;\mbox{tr}_C\{
[\tilde a+\tilde d][\frac{1}{6}s^2+\mbox{O}(\tilde cs^4)]+
\mbox{O}(\tilde b\tilde es^4)
+\mbox{O}([\tilde b^2+\tilde e^2]s^4)
\}\;\rangle_G
\;\;. \end{eqnarray}
Terms which are linear or cubic in $G_{\mu\nu}$ (and correspondingly in 
$e$) do not contribute  
here because of the Gaussian ensemble average over vacuum 
fields; the leading terms O($e^{\;0}$) are cancelled by the subtraction 
of ${\cal L}_0^-$, Eq.\,(\ref{LQCD}). Furthermore,  
using Eqs.\,(\ref{ad})--(\ref{cf}), we observe that the term which 
does not contain color fields, 
i.e. $\propto s^{-1}e^2[\tilde a+\tilde d]
\propto s^{-1}e^2F_{\mu\nu}F^{\mu\nu}$ in the integrand of Eq.\,(\ref{LQEDdiv}), 
presents the usual UV divergence which is absorbed by electromagnetic charge and field renormalization. Due to the ensemble average, 
in particular with $\langle G_{\alpha\beta}^aG_{\gamma\delta}^b\rangle_G
\propto\delta^{ab}(g_{\alpha\gamma}g_{\beta\delta}
-g_{\alpha\delta}g_{\beta\gamma})$, 
cf. the correlator (\ref{correlator}), the remaining finite 
terms $\propto s$ all are proportional to $e^2F_{\mu\nu}F^{\mu\nu}$ 
in the end, besides the appropriate power of the gluon condensate 
(\ref{condensate}). Therefore, they are absorbed by an additional 
finite renormalization.   
 
Following the previous analysis, we obtain the 
{\it renormalized} 
QCD-modified Euler-Heisenberg Lagrangian:
\begin{eqnarray} \label{GammaFin} \nonumber
{\cal L}_{EH}&=&-\frac{1}{4}F_{\mu\nu}F^{\mu\nu}
+{\cal L}_A^--{\cal L}_0^-
\\
\nonumber
&-&\frac{e^2}{1536\pi^2}
\int_0^\infty\frac{\md s}{s^2}\mbox{e}^{-im^2s}
\langle\;\mbox{tr}_C\{
\frac{1}{(\sqrt{\tilde c}\sin [s\sqrt{\tilde c}])^3}
\\
\nonumber
&\phantom .&
\left(
\cos [s\sqrt{\tilde c}](-3[\tilde b^2+\tilde e^2]
+12\tilde c[\tilde a+\tilde d]
+26s^2\tilde b\tilde c\tilde e
+11s^2\tilde c[\tilde b^2+\tilde e^2])
\right .
\\ 
\nonumber 
&\phantom .&\;\; 
+\cos [3s\sqrt{\tilde c}](3[\tilde b^2+\tilde e^2]
-12\tilde c[\tilde a+\tilde d]
-2s^2\tilde b\tilde c\tilde e
+s^2\tilde c[\tilde b^2+\tilde e^2])
\\
&\phantom .&\;\; 
\left .
-24s\sqrt{\tilde c}\sin [s\sqrt{\tilde c}]
(2\tilde c[\tilde a+\tilde d]
+\tilde b\tilde e 
\right) 
\}\;\rangle_G\;\;, \end{eqnarray}
where all charges and fields are the renormalized ones by now, in particular $\alpha\equiv e^2/4\pi\approx 1/137$, and where we implemented $\tilde c=\tilde f$ for the case of colormagnetic vacuum fields. Naturally, the zeroth order Lagrangian for the electromagnetic 
field appears here on the right-hand side. 

We remark that after renormalization and upon expansion of the effective action   
${\cal L}_{EH}$,  Eq.\,(\ref{GammaFin}), the QCD effects 
enter at O($e^4$) and higher, i.e. affecting the 
nonlinear effective interaction of electromagnetic fields.  
In the following section, we calculate the first nontrivial 
modification of the usual Euler-Heisenberg Lagrangian. 
Some comments concerning the nonperturbative content of our 
results will be made shortly. 

\subsection{Evaluation at O($e^4$)}

Here we expand the renormalized 
QCD-modified Euler-Heisenberg Lagrangian up to O($\alpha^2$), 
keeping all orders in the strong coupling. We consider as 
an instructive example the case of the stochastic   
colormagnetic vacuum fields $\vec B^a$ together with an (applied) external electric field $\vec E$. Thus Eqs.\,(\ref{ad})--(\ref{cf}) are 
replaced by: 
\begin{eqnarray} \label{ad1}
\tilde a&=&\tilde d=\frac{1}{2}F_{\mu\nu}F^{\mu\nu}=
-\vec E^2
\;\;, \\ \label{be1}
\tilde b&=&-\tilde e=-4ig\vec E\cdot\vec B^at^a 
\;\;, \\ \label{cf1} 
\tilde c&=&\tilde f=g^2\vec B\cdot\vec B
\;\;. \end{eqnarray}
Using these, 
we obtain ${\cal L}_{EH}$ at O($e^4$) from 
Eq.\,(\ref{GammaFin}),     
\begin{eqnarray} \label{e4int} \nonumber 
&\phantom .&{\cal L}_{EH}^{(4)}=
-\frac{1}{4}F_{\mu\nu}F^{\mu\nu}
\\ \nonumber 
&\phantom .&\; +\; 2\alpha^2\int_0^\infty\frac{\md x}{x^3}
\left\langle\mbox{tr}_C\left [\tilde c\;\mbox{e}^{-zx}
\right .\{ 
[{{-3{\tilde b^4}}\over {128{\tilde c^4}}} + 
       {{\tilde a{\tilde b^2}}\over {16{\tilde c^3}}} + 
       {{{\tilde a^2}}\over {8{\tilde c^2}}}]\;{x^2}
+[{{-{\tilde b^4}}\over {96{\tilde c^4}}} + 
     {{\tilde a{\tilde b^2}}\over {24{\tilde c^3}}}]\;{x^4}
\right . 
\\ \nonumber 
&\phantom .&\;\;\; 
+([{{-5{\tilde b^4}}\over {128{\tilde c^4}}} + 
        {{3\tilde a{\tilde b^2}}\over {16{\tilde c^3}}} - 
        {{{\tilde a^2}}\over {8{\tilde c^2}}}]\;x - 
     [{{{\tilde b^4}}\over {192{\tilde c^4}}} - 
        {{\tilde a{\tilde b^2}}\over {12{\tilde c^3}}} + 
        {{{\tilde a^2}}\over {4{\tilde c^2}}}]\;{x^3} - 
     {{{\tilde b^4}}\over {720{\tilde c^4}}}{x^5})\;
   \coth (x) 
\\ \nonumber 
&\phantom .&\;\;\; 
+([{{3{\tilde b^4}}\over 
            {128{\tilde c^4}}} - 
          {{\tilde a{\tilde b^2}}\over {16{\tilde c^3}}} - 
          {{{\tilde a^2}}\over {8{\tilde c^2}}}]\;{x^2}
          -[{{-{\tilde b^4}}\over {96{\tilde c^4}}} 
          +{{\tilde a{\tilde b^2}}\over {24{\tilde c^3}}}]\;{x^4})\;
        {\coth^2(x)}
\\
&\phantom .&\;\;\; 
\left .
+[{{{\tilde b^4}}\over {64{\tilde c^4}}} - 
     {{\tilde a{\tilde b^2}}\over {8{\tilde c^3}}} + 
     {{{\tilde a^2}}\over {4{\tilde c^2}}}]\;{x^3}
\left.   {\coth^3(x)} 
\}\right ]\right\rangle_G
\;\;, \end{eqnarray}
where $z\equiv m^2/\sqrt{g^2\vec B\cdot\vec B}$, and   
where again we rotated the contour of integration ($s\rightarrow -ix$), as compared to Eq.\,(\ref{GammaFin});
of course, the integral is finite due to the renormalization, despite the appearance of individually divergent 
contributions.

Using the value of the gluon condensate given in Eq.\,(\ref{condensate}) together with Eqs.\,(\ref{correlator}) and the approximate  
up- and down-quark masses, $m_u\approx 5\,\mbox{MeV}\approx
m_d/2$, the parameter 
$z$ of the integral in Eq.\,(\ref{e4int}) can be expected 
to fluctuate around a value of roughly $10^{-4}$ and 
$10^{-3}$, respectively. The corresponding number for 
strange quarks is at least two orders of magnitude larger;  
their (and heavier quark) contributions will be completely  
negligible in the following. Furthermore, for the relevant  range of small $z$-values, the integral is very strongly power dominated by the term in the integrand $\propto x^5/x^3$, in the region where $\coth (x)\approx 1$. 
Thus, we calculate:
\begin{eqnarray} \label{e4int1} \nonumber 
{\cal L}_{EH}^{(4)}&=&
-\frac{1}{4}F_{\mu\nu}F^{\mu\nu}
+2\alpha^2\int_0^\infty\md x
\left\langle\mbox{tr}_C\; [ 
{{{-\tilde b^4}}\over {720{\tilde c^3}}}\mbox{e}^{-zx}
]\; {x^2}\;
   \coth (x) 
\right\rangle_G 
\\ \nonumber 
&=&
\frac{1}{2}\vec E^2
-\frac{\alpha^2}{360}\sum_{\mbox{evs.}}\left\langle
\frac{\tilde b^4}{\tilde c^3}[\frac{1}{2}\zeta (3,z/2)
-2z^{-3}]
\right\rangle_G 
\\ &=&
\frac{1}{2}\vec E^2
-\frac{\alpha^2}{180}\sum_{\mbox{evs.}}\left\langle
\frac{\tilde b^4}{\tilde c^3z^3}[1+\mbox{O}(z^3)]
\right\rangle_G 
\;\;, \end{eqnarray}
where $\zeta$ denotes the Riemann zeta function in two 
arguments, and we employed formulae 3.551\,3. and 9.521\,1. 
of Ref.\,\cite{GR} in the second and third equality, respectively. Since the color matrices involved here are 
diagonal, as discussed, the sum 
over the eigenvalues of the resulting matrix replaces the color trace.   

Then, recalling that the covariantly constant $SU(3)$ colormagnetic vacuum 
fields are of the form 
$\vec B\equiv\frac{1}{2}(\vec B^3\lambda_3
+\vec B^8\lambda_8)$, in terms of the Gell-Mann matrices $\lambda_{3,8}$, we obtain:
\begin{eqnarray} \label{e4int2} \nonumber 
{\cal L}_{EH}^{(4)}&=&
\frac{1}{2}\vec E^2
-\frac{256\; g\;\alpha^2}{180\sqrt{3}\; m^6}
\left\langle [
\frac{(\vec E\cdot\vec B^8)^4}{\vert\vec B^8\vert^3}
+\sum_\pm
\frac{(\vec E\cdot(\sqrt{3}\vec B^3\pm\vec B^8)/2)^4} {\vert (\sqrt{3}\vec B^3\pm\vec B^8)/2\vert^3}
]\right\rangle_G 
\\  
&=&
\frac{1}{2}\vec E^2
-\frac{64\; g\;\alpha^2}{15\sqrt{3}\; m^6}
\left\langle 
\frac{(\vec E\cdot\vec B^8)^4}{\vert\vec B^8\vert^3}
\right\rangle_G
\;\;, \end{eqnarray}
where the last equality follows from the fact that all 
three contributions are equal. This can be shown by a 
suitable coordinate transformation in the space of the 
colormagnetic fields, the ensemble of which is 
determined by the Gaussian distribution of Eq.\,(\ref{distrib}). 

Using the distribution (\ref{distrib}) of the stochastic fields, 
we finally obtain from Eq.\,(\ref{e4int2}) the    
{\it QCD-modified} Euler-Heisenberg effective action for the case of external electric fields at O($\alpha^2$):
\begin{eqnarray} \label{EHaction} \nonumber  
\Gamma_{EH}^{(2)}&=&\int\md^4x\; [\;
\frac{1}{2}\vec E^2-\frac{2\alpha^2}{45}(\vec E^2)^2
\cdot\frac{1}{m_e^{\;4}}
\\ 
&\phantom .&-\frac{2\alpha^2}{45}(\vec E^2)^2
\sum_{i=u,d,\;\dots}\frac{64\sqrt{2\pi}}{5}
\frac{\langle
\frac{\alpha_s}{\pi}:G_{\mu\nu}^aG^{\mu\nu}_a:
\rangle^{1/2}}
{m_i^{\;2}}
(\frac{q_i}{m_i})^4\;]
\;\;, \end{eqnarray}
where $m_i$ and $q_i$ denote the quark masses and charges,  
respectively, and 
where we incorporated the usual QED term 
($\propto m_e^{-4}$) due to $e^+e^-$ vacuum 
polarization.\,\cite{Schw51,IZ} 

For  
$m_u\approx 5\,\mbox{MeV}\approx m_d/2$, $q_u=2/3=-2q_d$, 
and the gluon condensate value of Eq.\,(\ref{condensate}), the sum in the last term 
in Eq.\,(\ref{EHaction}) gives a numerical factor, 
\begin{equation} \label{sum} 
\sum_{i=u,d,\;\dots}\mbox{constant}\cdot q_i^{\;4}m_i^{-6}\approx 3.86\;(1+10^{-3}+\;\dots\;)\cdot\frac{1}{m_e^{\;4}}
\;\;, \end{equation} 
where the dots indicate the negligible 
contribution of the heavier quarks; we factored out 
$m_e^{-4}$ for comparison with the QED result. 

Clearly, the result of Eq.\,(\ref{EHaction}) 
is nonperturbative in the strong coupling $\alpha_s$ and 
strongly dominated by the lightest {\it u}-quark term. 
 
The experimentally interesting situation with crossed 
electric and magnetic external fields, related to the vacuum birefingence mentioned in Sec.\,1,\,\cite{BB70,EXP} 
can be studied along the same lines. Presently   
we presented only the simplest nontrivial 
case, in order to demonstrate that a sizeable  
interference between QCD and QED vacuum polarization     
appears to be possible, based on the present calculation.  
  
A thorough investigation of this effect needs to be 
performed which also takes the {\it space-time dependent}  stochastic vacuum fields into account, which are implied by the SVM description of QCD vacuum structure.\,\cite{SVM,DFK,LGT} In particular, 
we recall from Sec.\,1 that they are expected to fluctuate 
strongly on roughly the same length scale as the light quark loops, which we considered here. Whether this 
aspect of the nonperturbative stochastic fields tends to reduce the quark vacuum polarization considerably or not, still remains to be seen.     
    
\section{Discussion} 

We derived the modification of the Euler-Heisenberg 
effective action due to the nonperturbative QCD vacuum 
structure, which influences the vacuum polarization 
contribution of electrically and color charged 
quarks. We employed a simplified 
version of the stochastic vacuum model (SVM), which  
has been successful in describing 
various infrared aspects of QCD including confinement and bound state properties.\,\cite{SVM,DFK,LGT}  
 
In Sec.\,4.2 we obtained the light quark contribution to 
the effective action at O($\alpha^2$) for the case of 
external (macroscopic) electric fields acting simultaneously with stochastic colormagnetic vacuum 
fields. For simplicity, we assumed the QCD fields 
to be covariantly constant, such as in earlier 
background field calculations,\,\cite{Savv77,CY80} however, to be fluctuating in amplitude and (color) space 
direction. We found a sizeable contribution, of the 
same order of magnitude as the corresponding 
electron-positron term.  

The extension for the case of 
crossed electric and magnetic fields, such as employed in 
the experimental search for the usual QED vacuum birefringence effect,\,\cite{BB70,EXP} can be obtained 
in the same way.  
   
However, we discussed and want to stress here once 
more the fact that in order to obtain quantitatively reliable results, the inhomogeneous SVM fields necessarily have to be incorporated. Their correlation 
length is expected to interfere in a still unknown way 
with the length scale 
of the presumably dominant light quark vacuum polarization loops.

We conclude that a calculational scheme is 
necessary which allows to handle the important inhomogeneous field configurations.\,\cite{SVM,DFK} It is well known that applications of the Fock-Schwinger technique are limited to very special field configurations, such as the covariantly constant ones considered here.\,\cite{Schw51,IZ} 
 
Furthermore, our present study indicates that 
also the properties of the quark condensate, in external 
electromagnetic fields in particular, may deserve a fresh 
look when nonperturbative QCD features are implemented 
with the help of the stochastic vacuum model.
   
From the definition of the QCD effective action in the 
presence of external electromagnetic fields, Eq.\,(\ref{Gamma}) in Sec.\,2, it follows immediately 
that the quark condensate could be calculated nonperturbatively indeed:\,\cite{Schw51}  
\begin{equation} \label{qcondensate}
\langle\bar\psi\psi\rangle =-\frac{\md\Gamma_A}{\md m}
\;\;. \end{equation}   
Employing our Eqs.\,(\ref{LQCDstrong}) and 
(\ref{EHaction}), adding up the contributions, 
however, the result is  
quite unsatisfactory. This   
must be attributed to the fact that the homogeneous stochastic fields employed in a one-loop calculation here  
still present too crude an approximation for the relevant 
QCD vacuum properties. It will be very interesting to compare future results of an improved SVM calculation with other phenomenological models.\,\cite{chiralrest}  
Some related issues have already been discussed previously  
in a wider context.\,\cite{JR98} 

Finally, we mention that the fate of 
nonperturbative particle production poles contained in 
the effective Lagrangian (\ref{GammaFin}) of Sec.\,4.1 
after the stochastic averaging over vacuum fields also deserves further study.\,\cite{Schw51,IZ} 

\section*{Acknowledgments}
HTE wishes to thank A. Di\,Giacomo, H. G. Dosch, E. Ferreira, and T. Kodama for discussions and particularly A. Di\,Giacomo for his hospitality at the Dipartimento di Fisica (University of Pisa); the invitation by C. A. 
Bertulani to this stimulating workshop 
is gratefully acknowledged. This work was supported in part by CNPq-300758/97-9 (Brasil), 
PRONEX-41.96.0886.00, FAPESP-95/4635-0, 
and by a grant from the U.S. Department of
Energy, DE-FG03-95ER40937.

\end{document}